\documentclass[%
reprint,
showpacs,preprintnumbers,
amsmath,amssymb,
longbibliography
]{revtex4-2}

\usepackage{graphicx}
\usepackage{dcolumn}
\usepackage{bm}
\usepackage{colortbl}
\usepackage{lipsum}
\usepackage[unicode=true,colorlinks=true,citecolor=blue,urlcolor=blue]{hyperref}
\usepackage{braket}
\usepackage[normalem]{ulem}

\newcommand{\e}{\mathrm{e}}

\let\ifr\i
\renewcommand{\i}{{\rm i}}
\renewcommand{\d}{\mathrm d}
\renewcommand{\emph}{\textit}

\newcommand{\eps}{\varepsilon}
\newcommand{\kp}{$\bm k \cdot \bm p$}

\begin{document}

\title{Tunable linear polarization of interface excitons at lateral heterojunctions}

\author{M.~V.~Durnev}
\affiliation{Ioffe Institute, 194021 St. Petersburg, Russia}
\author{D.~S.~Smirnov}
\email[Electronic address: ]{smirnov@mail.ioffe.ru}
\affiliation{Ioffe Institute, 194021 St. Petersburg, Russia}

\begin{abstract}
 We develop a theory of polarized photoluminescence of interface excitons localized at lateral heterojunctions between transition metal dichalcogenide monolayers. We show that the circular selection rules governing interband optical transitions exactly at the band extrema are modified at finite wave vectors. The corresponding wave-vector-dependent corrections to the optical matrix elements result in a net linear polarization of excitonic photoluminescence. We identify two microscopic mechanisms responsible for linear polarization---trigonal warping of the electron and hole dispersions and the energy dependence of the effective masses. Their interplay controls both the magnitude and the angle of the emitted light polarization, with distinct dependences on the crystallographic orientation of the interface. Using a microscopic variational approach, we demonstrate that the degree of linear polarization can reach values exceeding 10\% in realistic heterostructures. Furthermore, due to the large built-in dipole moment of interface excitons, their optical response can be tuned by an external in-plane electric field, enabling control over the strength and direction of the polarization.
\end{abstract}

\maketitle

\section{Introduction}
\label{sec:intro}

Optical properties of transition metal dichalcogenide monolayers (TMD MLs) are predominantly governed by excitons with distinctive characteristics. Owing to reduced dielectric screening and relatively large effective masses of electrons and holes, excitons in TMD MLs exhibit large binding energies (on the order of fractions of an electronvolt), strong oscillator strengths, and small sizes~\cite{Wang:2018, Durnev:2018}. In analogy with the history of bulk semiconductors, early studies of TMD MLs have naturally evolved toward heterostructures, in particular, two-dimensional (2D) heterojunctions formed by vertically stacked monolayers~\cite{Geim:2013,Jin:2018,Castellanos-Gomez:2022}.

More recently, attention has been shifting toward one-dimensional lateral heterojunctions formed by different TMD MLs stitched together via covalent bonds along their edges~\cite{Li:2016,Avalos-Ovando:2019a,Castellanos-Gomez:2022}. Such lateral heterostructures can be synthesized by chemical vapor deposition using edge epitaxy~\cite{Duan:2014,Huang:2014,Ullah:2017,Sahoo:2018}. Notably, this approach enables the realization of atomically sharp interfaces with large lengths reaching tens of microns~\cite{Li:2015,Vandoolaeghe:2026}.

\begin{figure}[b]
\begin{center}
  \includegraphics[width=0.97\linewidth]{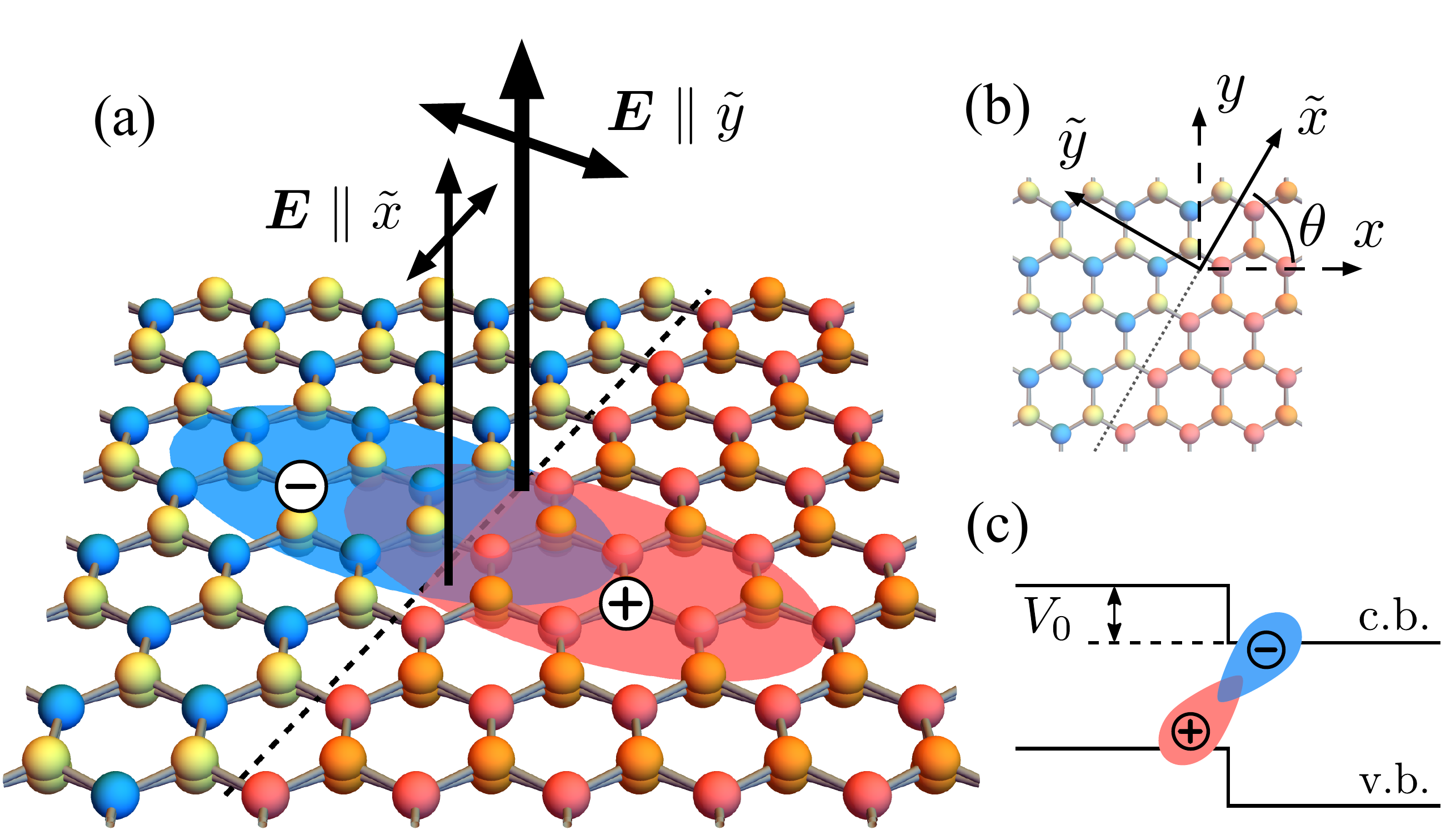}
 \end{center}
  \caption{\label{fig:system} Linearly polarized luminescence of interface excitons. (a) Sketch of an interface exciton localized at a lateral heterojunction. The intensity of the emitted radiation is different for electric field along and perpendicular to the interface resulting in the linear polarization of the excitonic photoluminescence.  (b) Lateral heterojunction oriented at an arbitrary angle $\theta$ with respect to the crystallographic axes $x$ and $y$. (c) Type-II band alignment of conduction and valence bands leads to the formation of a spatially indirect exciton.
  }
\end{figure}

The type-II band alignment between different TMD MLs~\cite{Kang:2013a,Guo:2016,Chu:2018,Herbig:2021,Sahoo:2026} promotes spatial separation of electrons and holes at the heterojunction, as shown in Fig.~\ref{fig:system}(a) and (c). At the same time, Coulomb attraction binds them together, resulting in the formation of a spatially indirect excitons localized at the interface~\cite{Lau:2018}. These interface excitons are characterized by an unusually large in-plane static dipole moment, reduced binding energy, and long radiative lifetimes~\cite{Rosati:2023,Lima:2023,Durnev:2025,Rosati:2025}. Experimentally, they manifest, for instance, as low-energy peaks in photoluminescence (PL) spectra, reflecting the ground state of excitons in lateral heterostructures~\cite{Rosati:2023,Yuan:2023,Vandoolaeghe:2026}. Overall, the study of interface excitons remains at an early stage, and fundamentally new effects specific to these states are yet to be explored.

In particular, the fine structure and polarization properties of interface excitons remain scarcely studied. In this work, we revisit the optical selection rules in TMD MLs and demonstrate that at finite wave vectors in the vicinity of the band extrema ($K$-points of the Brillouin zone), interband optical transitions become elliptically polarized. The admixture of linear polarization to circularly polarized optical transitions is governed by two mechanisms of distinct symmetry---trigonal warping of electron and hole spectra and the dependence of the effective masses on energy. For the localized excitons, such a modification of the selection rules leads to intrinsic linear polarization of their net PL. 

We show that the linear polarization is especially pronounced for interface excitons, as their PL is governed by the narrow region of electron-hole overlap, which is contributed mainly by the states with large electron and hole wave vectors. We derive general expressions for the Stokes parameters of the exciton emission and analyze their dependence on the crystallographic orientation of the lateral heterojunction for the interface excitons. Furthermore, by presenting microscopic calculations of the energy spectra and degree of intrinsic linear polarization of interface excitons in realistic lateral heterostructures, we demonstrate that both the magnitude and the orientation of the linear polarization can be tuned by an in-plane electric field applied across the heterojunction.

\section{Optical selection rules revisited}
\label{bulk_opt}

The interband optical transitions from the valence to the conduction band are described by the following matrix elements:
 \begin{equation}
 \label{Mcv}
M_{cv}(\bm q) = -\frac{e}{c} \mathcal A_0\, \bm v_{cv} (\bm q) \cdot \bm e\:,
\end{equation}
where $\bm v_{cv}$ is the interband matrix element of the velocity operator, $ \bm{\mathcal A}(t) = \mathcal A_0 (\bm e\, \e^{-\i \omega t} + \bm e^* \e^{\i \omega t})$ is the vector potential of the monochromatic field with the frequency $\omega$, $\bm e$ is the polarization vector, and $\bm q = (q_x, q_y)$ is the electron wave vector. The $\bm v_{cv}$ matrix element can be found from a particular Hamiltonian describing electron states. Optical properties of TMD monolayers are determined by  direct transitions between the bands edges located at $\bm K_\pm$ points of the Brillouin zone. We denote the velocity matrix elements in the corresponding valleys as $\bm v_{cv}^\pm(\bm k)$, where $\bm k$ is the wave vector measured from $\bm K_\pm$.

Exactly at the absorption edge, the transitions are circularly polarized with the opposite helicity in $K_+$ and $K_-$ valleys: $\bm v_{cv}^{(\pm)}(0) \cdot \bm e \propto e_\mp$, where $e_\pm = e_x \pm \i e_y$. However, at finite $\bm k$ the velocity matrix elements contain corrections, which can be found by expanding $\bm v_{cv} (\bm k)$ in a series in $\bm k$. The symmetry analysis in the $C_{3h}$ point group of the $K_\pm$ valleys shows that such an expansion (up to quadratic in $\bm k$ terms) has the form
\begin{eqnarray}
\label{vcv_symmetry}
\bm v_{cv}^{(+)} \cdot \bm e &=& \frac{\gamma_3}{\hbar} \left[e_- (1 - \alpha k^2) - A k_+e_+  - \beta k_- (\bm k \cdot \bm e) \right]\,,  \\
\bm v_{cv}^{(-)} \cdot \bm e &=& \frac{\gamma_3}{\hbar} \left[-e_+ (1 - \alpha k^2) - A k_-e_-  + \beta k_+ (\bm k \cdot \bm e) \right]\,, \nonumber
\end{eqnarray}
where $\gamma_3$, $A$, $\alpha$ and $\beta$ are real parameters and $k_\pm = k_x \pm \i k_y$ with the crystallographic axes $x$ and $y$ shown in Fig.~\ref{fig:system}(b). In Eqs.~\eqref{vcv_symmetry} we took into account that the two valleys are related by the time reversal symmetry and reflection in the $(yz)$ plane. Parameter $\alpha$ leads to suppression of the transitions rates and does not affect their polarization to the first order, therefore we will neglect it. Parameter $A$ reflects the trigonal symmetry of the valleys, which dictates that $e_-$ and $k_+e_+$ transform according to the same irreducible representation~\cite{Kormanyos:2013,Kormanyos:2015,Glazov:2017}. By contrast, the last term proportional to $\beta$ is isotropic since it transforms as $e_-$ under rotation by any angle around the $z$-axis.

Equations~\eqref{vcv_symmetry} show that the optical transitions in TMD monolayers are purely circularly polarized only at the corners of the Brillouin zone, i.e. at $\bm k=0$, whereas at finite $\bm k$ the transitions become elliptically polarized.

The values of the coefficients, which describe modification of the selection rules can be obtained from the atomistic  calculations. As an example, we adopted the tight-binding (TB) model from Ref.~\cite{Rybkovskiy:2017aa} for a monolayer MoS$_2$. It gives the parameters $\alpha = 0.54$~\AA$^2$, $A = 0.17$~\AA~and $\beta = 8.98$~\AA$^2$. They are quite small; strong deviations from the circular selection rules require $k\sim3$~nm$^{-1}$ ($|\bm K_\pm|\approx 13$~nm$^{-1}$). At the same time, the terms proportional to $A$ and $\beta$ become comparable at $k\sim0.2$~nm$^{-1}$ already. Another implemented TB parametrization results in similar values of the parameters, see App.~\ref{app:TB}.

The origin of the $A$ and $\beta$ terms in Eq.~\eqref{vcv_symmetry} can be illuminated in the framework of the two-band \kp-model which includes both linear and quadratic in $\bm k$ off-diagonal elements. The corresponding Hamiltonian in the $K_+$-valley has the form~\cite{Kormanyos:2013}
\begin{equation}
\label{H0}
\mathcal H^{(+)} = \left[ 
\begin{array}{cc}
E_g & \gamma_3 k_- - \frac12 A \gamma_3 k_+^2 \\
\gamma_3 k_+ - \frac12 A \gamma_3 k_-^2 & 0
\end{array}
\right]\:,
\end{equation}
where $E_{g}$ is the energy gap. The Hamiltonian $\mathcal H^{(-)}$ for the $K_-$-valley is obtained by replacing $k_\pm$ with $- k_\mp$. The velocity matrix elements are calculated as $\psi_c^\dag \bm v \psi_v$, where $\psi_{c/v}$ are the eigen columns of $\mathcal H^{(\pm)}$ and $\bm v = \hbar^{-1} d \mathcal H^{(\pm)}/d \bm k$. Up to the second order in $\bm k$ these matrix elements agree with Eq.~\eqref{vcv_symmetry} with $\alpha = 0$ and $\beta = 2 \gamma_3^2/E_g^2$.

At the same time, the energy dispersion of the conduction and valence bands in the model~\eqref{H0} up to $k^4$ reads
\begin{eqnarray}
E_c^{(\nu)}(\bm k) &=& E_g + \frac{\gamma_3^2}{E_g} \left(k^2-\frac{\beta}{2}k^4\right) - \nu \frac{\gamma_3^2}{E_g} A k^3\cos3\varphi\:, \nonumber \\
E_v^{(\nu)}(\bm k) &=& - \frac{\gamma_3^2}{E_g}\left(k^2-\frac{\beta}{2}k^4\right) + \nu \frac{\gamma_3^2}{E_g} A k^3\cos3\varphi\:, 
\end{eqnarray}
where $\varphi$ is the angle between $\bm k$ and the $x$ axis, $\nu = \pm 1$ is the valley index, and we neglect the terms proportional to $A^2$. These expressions clearly show that $A$ causes the trigonal warping of the isoenergetic contours, whereas $\beta$ is responsible for the deviation of the isotropic part of the dispersion from the parabolic one (dependence of the effective mass on energy). Microscopic expression for $A$ can also be obtained from the extended 4-band \kp-model, which takes into account additional conduction and valence bands~\cite{Glazov:2017, Durnev:2018}.

\begin{figure}[t!]
\begin{center}
  \includegraphics[width=0.97\linewidth]{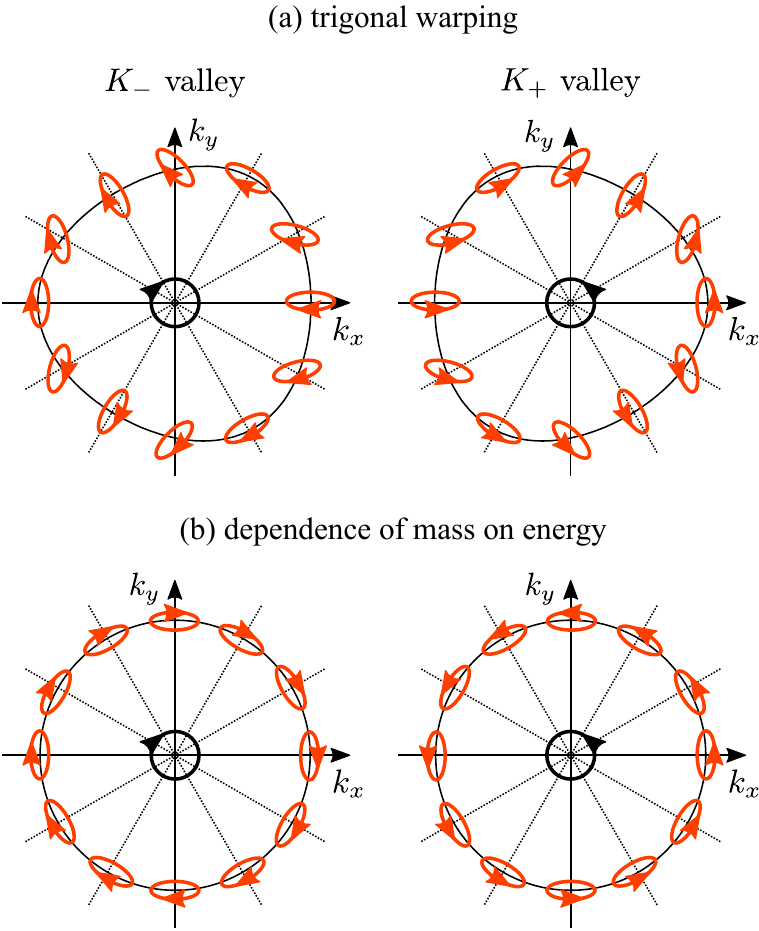}
 \end{center}
  \caption{\label{fig:selection_rules} The scheme of polarization of optical transitions in the $K_+$ and $K_-$ valleys of bulk TMD monolayers. The transitions occur in circular polarization strictly at $\bm k = 0$, whereas at $\bm k \neq 0$ the polarization becomes elliptical, see Eq.~\eqref{select_rules}. Polarization (a) due to the trigonal warping and (b) due to the dependence of mass on energy.
  }
\end{figure}

The rates of optical transitions in the $\bm K_\pm$ valleys are found from Eqs.~\eqref{Mcv} and \eqref{vcv_symmetry}. They are determined by $|\bm v_{cv}^{(\pm)} \cdot \bm e|^2$ which is proportional to 
\begin{align}
\label{select_rules}
|\bm v_{cv}^{(\pm)} \cdot \bm e|^2 &\propto (1 \pm P_c)(1 - 2\alpha k^2 - \beta k^2) \\ 
\mp 2A k_x P_l &\pm 2A k_y P_{l'} - \beta(k_x^2 - k_y^2) P_l - 2\beta k_x k_y P_{l'} \nonumber\:,
\end{align}
where $P_l = |e_x|^2 - |e_y|^2$, $P_{l'} = e_x e_y^* + e_x^* e_y$ and $P_c = \i(e_x e_y^* - e_x^* e_y)$ are the Stokes parameters, which describe horizontal/vertical, diagonal/anti-diagonal, and circular polarizations in the $x,y$ axes, respectively. These optical selection rules are illustrated in Fig.~\ref{fig:selection_rules} separately for the mechanisms related to the trigonal warping and dependence of mass on energy. One can see that the admixture of linear polarization described by the terms in the second line of Eq.~\eqref{select_rules} leads to the elliptically polarized optical transitions in each valley at finite $\bm k$.
The orientation of ellipses depends on $\bm k$ and is different for the terms related to the trigonal warping and the dependence of mass on energy. Note that the same selection rules are responsible for the optically induced valley currents and optical alignment of electron momenta in 2D Dirac materials~\cite{Golub:2011,Entin:2021, Durnev:2021vj, Saroka:2022}.

Modification of the optical selection rules is a general effect, which can be observed in bulk TMD monolayers in different conditions, for example for the localized excitons. We show below that its manifestations are particularly pronounced for the lateral heterostructures.

\section{Polarized luminescence of interface excitons}
\label{sec:PL}

  The exciton envelope wave function $\psi(\bm r_e,\bm r_h)$ in a given valley $\nu$  depends on the electron and hole 2D coordinates, $\bm r_e$ and $\bm r_h$, respectively. Matrix elements of optical transitions involving excitons are calculated by averaging the free electron-hole transitions as 
  \begin{equation}
  M^\nu=\int\tilde\psi^*(\bm k,-\bm k)M_{cv}^\nu(\bm k)\frac{\d\bm k}{(2\pi)^2},
\end{equation}
  where
  \begin{equation}
    \label{Fourier}
  \tilde\psi(\bm k_e, \bm k_h)=\int \psi(\bm r_e,\bm r_h)\e^{-\i\bm k_e\bm r_e-\i\bm k_h\bm r_h}\d\bm r_e\d\bm r_h
\end{equation}
is the exciton wave function in the momentum space. Here, $M_{cv}^\nu(\bm k)$ is given by Eq.~\eqref{Mcv} with the velocity matrix elements in the corresponding valley~\eqref{vcv_symmetry} and we took into account that the exciton wave function is written in the electron-hole representation which is related to the electron representation by the time reversal for the states of the valence band.

In practice, it is more convenient to use the coordinate representation which yields
  \begin{equation}
    \label{Mpm_r}
    M^\nu=\int \delta(\bm r_e-\bm r_h)M_{cv}^\nu\left(-\i\frac{\partial}{\partial\bm r_h}\right)\psi^*(\bm r_e,\bm r_h)\d\bm r_e \d\bm r_h\:.
  \end{equation}
 The Stokes parameters of light emitted by excitons in the valley $\nu$ are related to $M^\nu$ as:
  \begin{align}
    \label{Stokes}
    P_l^\nu &= \frac{|M^\nu_x|^2 - |M^\nu_y|^2}{|M^\nu_x|^2 + |M^\nu_y|^2}\:,~P_{l'}^\nu = \frac{|M^\nu_{x'}|^2 - |M^\nu_{y'}|^2}{|M^\nu_{x'}|^2 + |M^\nu_{y'}|^2}\:, \nonumber\\
    P_{c}^\nu &= \frac{|M^\nu_{\sigma+}|^2 - |M^\nu_{\sigma-}|^2}{|M^\nu_{\sigma+}|^2 + |M^\nu_{\sigma-}|^2}\:,
  \end{align}
  where the subscript denotes the corresponding light polarization vector~\footnote{Explicitly, $\bm e_{x'/y'} = (\bm e_x\pm\bm e_y)/\sqrt{2}$ and $\bm e_{\sigma\pm} = (\bm e_x \pm \i \bm e_y)/\sqrt{2}$.}.

Equations~\eqref{Mpm_r} and \eqref{Stokes} are general and can be applied to any excitonic states. Now let us consider interface excitons localized at the most natural zigzag interface directed along the $x$ axis. The wave function of an exciton (with the zero center-of-mass wave vector along the interface) has the form $\psi(\bm r_e, \bm r_h) = \Phi(x,y_e,y_h)$, where $x=x_e-x_h$. Neglecting the corrections to the velocity matrix elements in Eq.~\eqref{vcv_symmetry}, the exciton photoluminescence is circularly polarized, as usual, so that $P_c^\nu = \nu$ for excitons in the two valleys and the linear polarization is absent. The PL intensity is determined by the overlap integral
  \begin{equation}
    \kappa_0=\int\Phi(0,y,y)\d y.
  \end{equation}
Due to the type-II band alignment, the overlap is related to the tunnelling of electron and hole under the potential barrier and, hence, is small, see Fig.~\ref{fig:system}(c).
  
Corrections to the velocity matrix elements in Eq.~\eqref{vcv_symmetry} result in linear polarization of exciton PL. The corresponding nonzero Stokes parameter in the first order in $A$ and $\beta$ reads
  \begin{equation}
    P_l^\nu=A\kappa_1-\beta\kappa_2,
  \end{equation}
  where
  \begin{align}
    \label{kappas}
    \kappa_1 &= -\frac{2}{\kappa_0}\int \frac{\partial}{\partial y} \Phi(0,y',y)\Big|_{y'=y} \d y \:, \\
        \kappa_2 &= \frac{1}{\kappa_0}\int \left(\frac{\partial^2}{\partial y^2}-\frac{\partial^2}{\partial x^2} \right) \Phi(x,y',y)\Big|_{x=0,y'=y} \d y \nonumber \:.
   \end{align}
 The polarization is directed along or perpendicular to the heterojunction depending of the sign of $P_l$, in agreement with the $C_{2v}$ symmetry of the zigzag interface which includes the $(yz)$ mirror reflection plane.

In a more general case of an interface making an angle $\theta$ with the $x$ axis, as shown in Fig.~\ref{fig:system}(b), the calculation of the Stokes parameters is analogous. Again, up to the first order in $A$ and $\beta$ we obtain $\tilde P_c^\nu=\nu$ and
  \begin{eqnarray}
    \label{Xpol}
    \tilde P_l^\nu &=& A \kappa_1 \cos 3 \theta  - \beta\kappa_2, \nonumber \\
    \tilde P_{l'}^\nu  &=& -A \kappa_1 \sin 3 \theta \:,
  \end{eqnarray}
where tilde denotes the Stokes parameters in the interface-related axes $(\tilde x,\,\tilde y)$, see Fig.~\ref{fig:system}(b). Equations~\eqref{Xpol} reflect the nature of the two mechanisms of linear polarization. The mechanism related to the energy-dependent mass is isotropic, hence the corresponding term in Eq.~\eqref{Xpol}, $\beta\kappa_2$, does not depend on the interface angle $\theta$. The resulting optical transitions are polarized either along or perpindicular to the interface (depending on the sign of $\beta \kappa_2$), hence this mechanism contributes to $\tilde P_l^\nu$ only.

\begin{figure}
\begin{center}
  \includegraphics[width=0.9\linewidth]{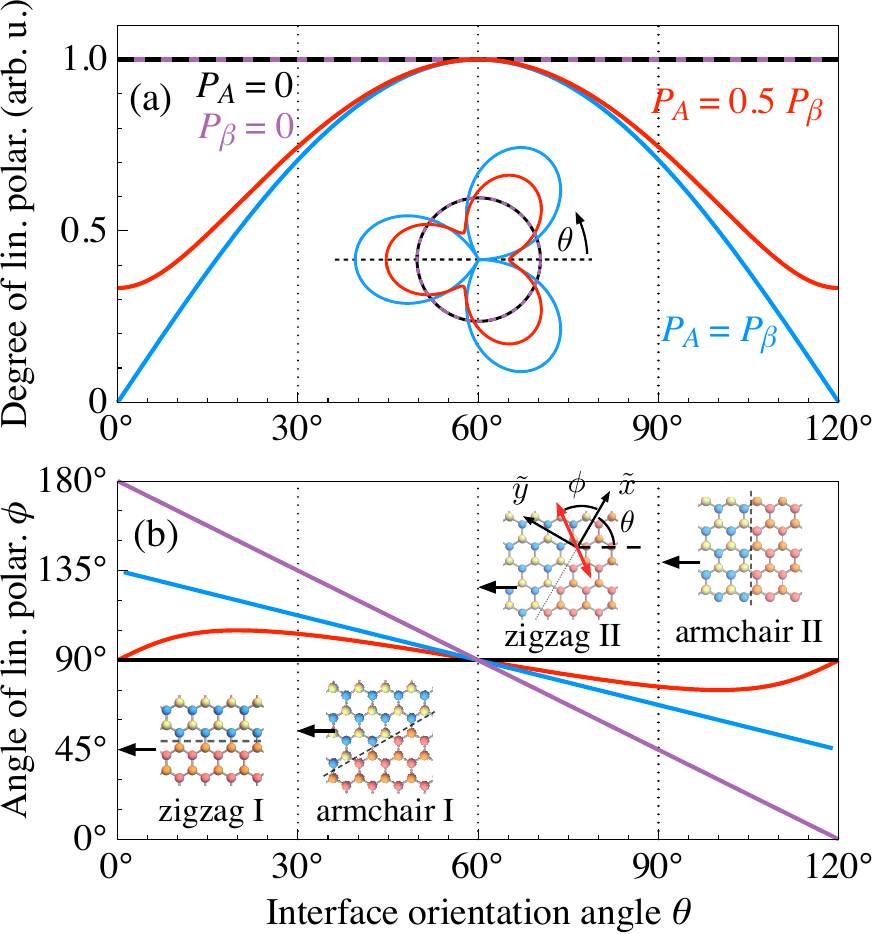}
 \end{center}
 \caption{\label{fig:Ptheta}  Linear polarization of the interface excitons as a function of the interface orientation. (a) Degree of linear polarization $P_{\rm lin}$ for different ratios between the two contributions $P_A = A \kappa_1$ and $P_\beta = \beta \kappa_2$ in Eq.~\eqref{Xpol}. The inset shows the same in the form of a polar plot. (b) Direction of linear polarization with respect to the interface. The insets depict the heterostructures corresponding to the two different zigzag ($\theta = 0,\,60^\circ$) and armchair ($\theta = 30^\circ,\,90^\circ$) interfaces.
  }
\end{figure}

By contrast, the trigonal warping mechanism leads to linear polarization with the degree varying as the third angular harmonic of $\theta$. This mechanism contributes to both Stokes parameters $\tilde P_l^\nu$ and $\tilde P_{l'}^\nu$.
Note that the angles $\theta=n\pi/3$ with integer $n$ correspond to the zigzag interfaces, whereas armchair interfaces are realized at $\theta=n\pi/3+\pi/6$. However, the trigonal warping contributions ($\propto A\kappa_1$) change sign every $\pi/3$ reflecting that there are, in fact, two types of zigzag and armchair interfaces, denoted by I and II in the insets in Fig.~\ref{fig:Ptheta}(b). Zigzag interfaces of type I and II are distinguished by the atoms closest to the interface, whereas armchair interfaces of different types have different arrangement of atoms in the $\tilde x$ direction. The interfaces that differ by $\Delta \theta = 2\pi/3$ are equivalent.

Importantly, as follows from Eq.~\eqref{Xpol}, the linear polarization is the same for excitons in both valleys. It means that the total photoluminescence from an equally populated ensemble of $K_+$ and $K_-$ excitons is linearly polarized. There is also an inverse effect: the absorbance at the resonance frequency of the interface exciton depends on the direction of linear polarization of light, i.e. the lateral heterojunction exhibits linear dichroism at exciton resonance.

For a given ratio between the two mechanisms of linear polarization, the degree of linear polarization
  \begin{equation}
    P_{\rm lin} = \sqrt{\left(\tilde{P}_l^\nu\right)^2 + \left(\tilde{P}_{l'}^\nu\right)^2}
  \end{equation}
  depends on the direction of the interface $\theta$. This is illustrated in Fig.~\ref{fig:Ptheta}(a) for different ratios between $P_A = A \kappa_1$ and $P_\beta = \beta \kappa_2$ (we assume $P_A > 0$ and $P_\beta > 0$). In general, the dependence is $2\pi/3$-periodic, as shown in the inset, which reflects the trigonal symmetry of the lattice.
  
The direction of linear polarization makes the angle $\phi = 1/2 \arctan (\tilde{P}_{l'}^\nu/\tilde{P}_{l}^\nu)$ with the interface axis $\tilde x$. This angle is even more sensitive to the interface orientation, as shown in Fig.~\ref{fig:Ptheta}(b). For $P_A = 0$, we have $\phi=\pi/2$ in line with the above discussion. For $P_\beta=0$, one obtains $\phi = \pi-3\theta/2$, so that optical transitions are polarized along the zigzag I interface and perpendicular to the zigzag II interface, whereas for the armchair interfaces one has $\phi = \mp \pi/4$. In general case, the dependences of $P_{\rm lin}$ and $\phi$ on $\theta$ are nontrivial and allow one to clearly distinguish between the different types of the armchair and zigzag interfaces. Note that in the double heterostructure of the A/B/A form, the two parallel interfaces differ by $\Delta \theta = \pi$, and hence have different types.

According to Eq.~\eqref{kappas}, the absolute value of the degree of linear polarization depends on the specific form of the exciton wave function. The dominant contribution to $\kappa_1$ and $\kappa_2$ is given by a small region where electron and hole wave functions overlap. However, the wave functions change abruptly in this region, which means that the typical wave vectors determining the polarization degree are large. This allows one to expect a sizable polarization despite the moderate values of $A$ and $\beta$ and the small overlap between electron and hole wave functions. This suggestion is corroborated by microscopic modelling in the next section.

\section{Tunable interface excitons}
\label{sec:results}

\subsection{Microscopic model}

To calculate the wave functions and optical properties of interface excitons, we adopt the variational method developed in Ref.~\cite{Durnev:2025}. Let us briefly summarize it here.

The interface-exciton Hamiltonian in the effective mass approximation reads~\cite{Berkelbach:2013,Chernikov2014,Lau:2018,Durnev:2025}:
\begin{equation}
  \label{eq:H_eh}
  \mathcal H_X = \frac{\hbar^2 k_e^2}{2m_c}+\frac{\hbar^2 k_h^2}{2m_v}+V_e(\bm r_e)+V_h(\bm r_h)+V_{\rm RK}(|\bm r_e-\bm r_h|).
\end{equation}
Here, $m_{c,v}$ are the electron and hole effective masses, $V_{e,h}$ are the conduction- and valence-band energy profiles (which can include an external electrostatic potential), $V_{\rm RK}(r)=-(\pi e^2)/(2r_0')\left[H_0(r/r_0')-Y_0(r/r_0')\right]$ is the Rytova-Keldysh potential~\cite{Rytova,Keldysh1979} with $H_0$ and $Y_0$ being the Struve and Neumann functions, respectively, $e>0$ is the elementary charge, $r_0' = r_0/\epsilon$, where $r_0$ is the screening length of a monolayer in vacuum and $\epsilon=(\epsilon_1+\epsilon_2)/2$ is the average dielectric constant of the media above and below the monolayer. The conduction- and valence-bands potentials (in the electron-hole representation) are taken in the form corresponding to an atomically sharp interface:
\begin{equation}
\label{VeVh}
  V_{e,h}(\bm r) = V^{(e,h)}_0\Theta(\mp\tilde{y}) \pm e F \tilde y,
\end{equation}
where $V^{(e,h)}_0$ are the band offsets and $\Theta(\tilde y)$ is the Heaviside step function [we recall that $\tilde y$ is perpendicular to an arbitrary oriented interface, see Fig.~\ref{fig:system}(b)]. Importantly, the last term in Eq.~\eqref{VeVh} allows us to account for an external electric field $F$ applied perpendicularly to the interface. Note that this effective mass approach is equally valid for any orientation of an interface.

The Schr\"odinger equation is solved using the variational approach. Since the discontinuity of the energy gaps is much smaller then the typical band offsets, we take $V^{(e)}_0 = V^{(h)}_0 = V_0$, while keeping all the other parameters such as $E_g$, $\gamma_3$, $A$, $\beta$, $r_0$ and $m_{c,v}$ the same at the both sides of the interface. Setting additionally $m_c=m_v$, the trial wave function can be chosen in a symmetric form~\cite{Durnev:2025}:
\begin{equation}
  \label{PhiX}
  \Phi(\tilde x, \tilde y_e, \tilde y_h) = \left(\frac{2}{\pi c^2}\right)^{1/4}\e^{- \tilde x^2/c^2} f(\tilde y_e) f(-\tilde y_h)\:,
\end{equation}
where
\begin{multline}
  \label{fx}
  f(\tilde y)=\frac{2}{(a+b)\sqrt{2a+b}}\left[\left(\tilde y+\frac{a}{b} \tilde y+a\right)\e^{-\tilde y/b}\Theta(\tilde y)\right.\\\left.+a\e^{\tilde y/a}\Theta(-\tilde y)\right]\:,
\end{multline}
and $a$, $b$, and $c$ are the variational parameters.

\begin{figure}
\begin{center}
  \includegraphics[width=\linewidth]{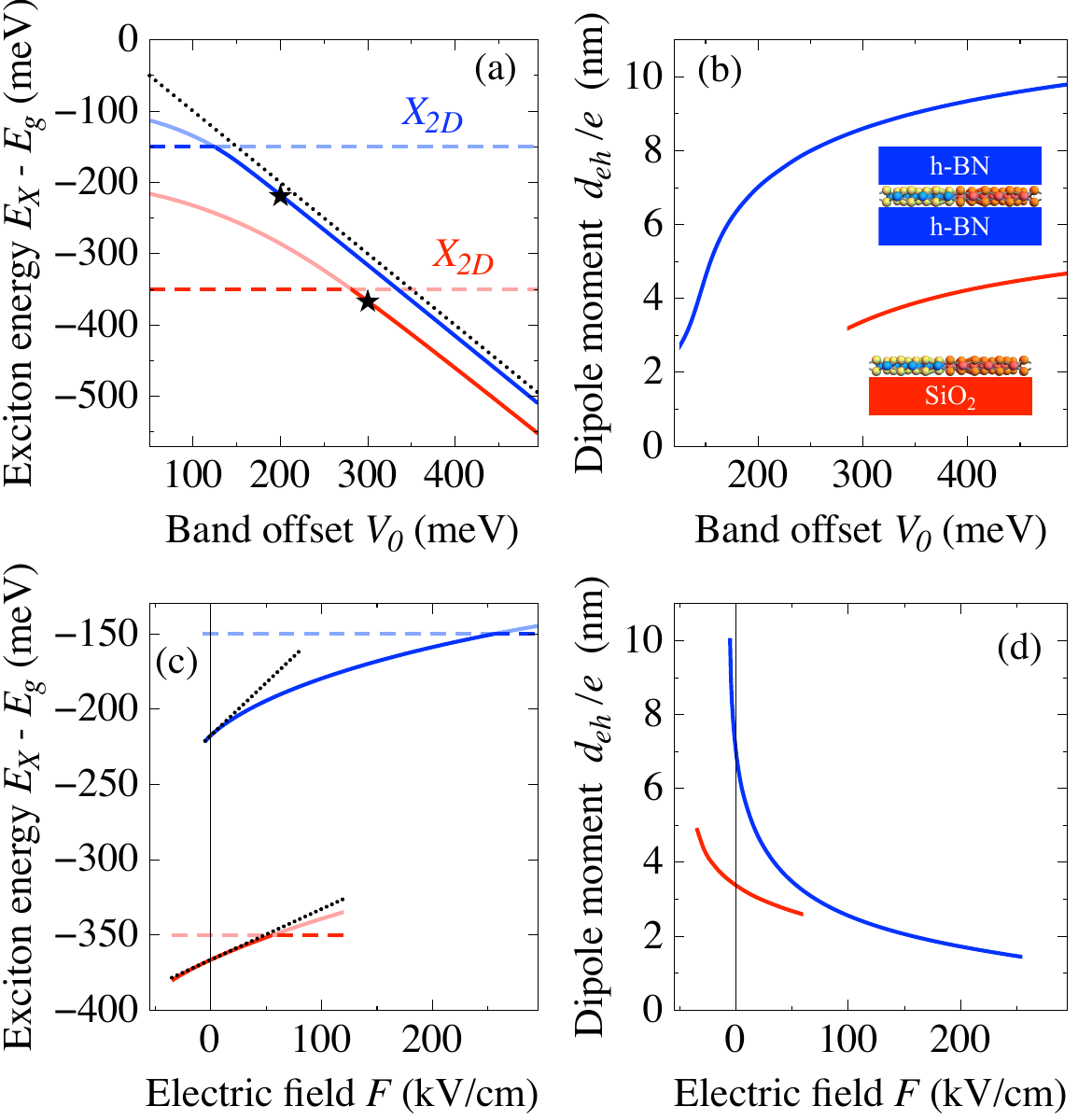}
 \end{center}
 \caption{\label{fig:Xspectra}   Tunable interface excitons in lateral heterostructures on a SiO$_2$ substrate (red lines) and encapsulated in h-BN (blue lines). (a) Exciton energy (solid lines) counted from the band gap as a function of the band offset. Dashed lines show the energies of ``bulk'' 2D excitons, black dotted line shows the $-V_0$ function. (b) Static dipole moment as a function of the band offset. The inset schematically depicts the studied structures. (c) Exciton energy as a function of electric field applied perpendicularly to the interface. The curves are calculated using $V_0 = 200$~meV for the encapsulated structure and $V_0 = 300$~meV for the structure on a substrate [shown by stars in panel (a)]. The dotted black lines show linear approximation with the slope determined by $d_{eh}$ at $F = 0$. (d) Static dipole moment as a function of electric field.
 }
\end{figure}

In our calculations we use $m_c = m_v = 0.4m_0$, $r_0 = 4.8$~nm and consider two different structures: One is a lateral heterostructure supported by a SiO$_2$ substrate with static and high-frequency dielectric constants $\epsilon_0 = 2.45$ and $\epsilon_\infty = 1.55$, respectively; and another structure is encapsulated in hexagonal boron nitride (h-BN) with $\epsilon_0 = 4.9$ and $\epsilon_\infty = 4.45$~\cite{Durnev:2025}.

Energy minimization allows one to calculate the energy of an interface exciton. It is shown in Fig.~\ref{fig:Xspectra}(a) as a function of the band offset. The interface exciton becomes a ground state when its energy is lower than that of the bulk 2D exciton (shown by the dashed lines). It occurs for the band offsets exceeding a certain threshold, of about $120$~meV and $280$~meV for the encapsulated and supported structures, respectively. We further calculate properties of interface excitons in these regions.

\subsection{Results and discussion}

We begin the analysis by considering an exciton static dipole moment. For the wave function~\eqref{PhiX} it is given by $d_{eh} = e(3b^2+4ab-2a^2)/(b+2a)$ and is directed opposite to the $\tilde y$ axis. In the feasible limit of weak penetration of the wave functions under barriers, $a\ll b$, we obtain $d_{eh}\approx 3eb$. The dipole moment is shown in Figs.~\ref{fig:Xspectra}(b) and (d). One can see that increase of the band offset leads to the increase of the dipole moment, as expected, Fig.~\ref{fig:Xspectra}(b). $d_{eh}$ can also be tuned by the static electric field applied perpendicularly to the interface, Fig.~\ref{fig:Xspectra}(d). In the chosen range of $F$, the dipole moment varies from about $2$ to $10$~$e\cdot$nm for the encapsulated structure.

Figure~\ref{fig:Xspectra}(c) shows the exciton energy as a function of electric field. In the range of small fields, one can see a linear Stark shift with the slope determined by the dipole moment at $F = 0$, as shown by the dotted lines. However, since $d_{eh}$ changes significantly with the field, the overall dependence of the exciton energy on $F$ is nonlinear. Negative electric fields tend to separate electron and hole in the exciton. At a certain value of $F$ the local minimum of the energy in the variational calculation disappears, which corresponds to the exciton dissociation. The critical values of the field are about $-35$~kV/cm and $-5$~kV/cm for the supported and encapsulated structures, respectively. Positive electric fields push electron and hole towards each other so that the dipole moment decreases and the exciton energy increases. When the energy reaches the bulk exciton energy [shown by the dashed lines in Fig.~\ref{fig:Xspectra}(c)], the electron/hole tunnels through the potential barrier and a spatially direct exciton forms. The corresponding fields are quite large, of about $55$ and $235$~kV/cm for the supported and encapsulated structures, respectively.

\begin{figure}
\begin{center}
  \includegraphics[width=\linewidth]{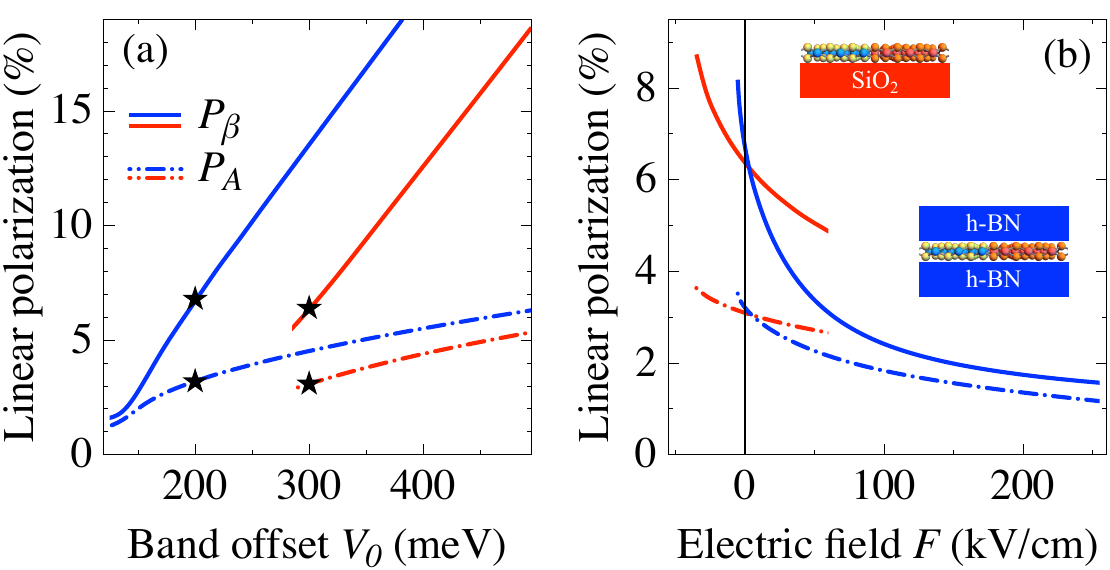}
 \end{center}
 \caption{\label{fig:X_Pl} Linear polarization of the interface exciton emission. Solid and dashed lines show the contributions related to the energy dependence of mass ($P_\beta$) and the trigonal warping ($P_A$), respectively. Blue and red lines correspond to encapsulated and supported structures. Dependences of $P_\beta$ and $P_A$ (a) on the band offset and (b) on the electric field are shown. The curves in (b) are plotted for the band offsets indicated by the stars in (a).
 }
\end{figure}

Linear polarization of the interface exciton emission is determined by the parameters $\kappa_{1,2}$ in Eqs.~\eqref{kappas} and~\eqref{Xpol}. For the exciton wave function~\eqref{PhiX}, we obtain
\begin{equation}
\kappa_1 = \frac{3b - a}{2 a b}\:,\qquad \kappa_{2} = \frac{b-a}{2 a^2 b}+\frac{2}{c^2}\:.
\end{equation}
The contributions to the exciton Stokes parameters related to the energy dependence of mass ($P_\beta$) and the trigonal warping ($P_A$) are shown in Fig.~\ref{fig:X_Pl} as functions of the band offset and electric field. Generally, one can see that the former mechanism dominates, and the polarization is larger for the encapsulated structures. Depending on the parameters, the absolute value of polarization can be sizeable and exceed $10$\%.
 
We recall that the Stokes parameters of the exciton emission, Eq.~\eqref{Xpol}, depend significantly on the ratio between the two contributions $P_A$ and $P_\beta$. Changing this ratio with electric field, see Fig.~\ref{fig:X_Pl}(b), allows one to manipulate not only the degree of polarization, but also its direction, as shown in Fig.~\ref{fig:Ptheta}.

The above calculations rely on the small-$\bm k$ expansion in Eq.~\eqref{vcv_symmetry} for the optical matrix elements and, strictly speaking, can be applied only for $P_A \ll 1$ and $P_\beta \ll 1$. For comparison, we have also performed calculations of the linear polarization using the tight-binding model, see App.~\ref{app:TB} for details. These calculations show that for polarizations up to 10\%, the difference between two approaches remains below 1\%.

\section{Summary}
\label{sec:summary}

To summarize, we have shown that the light emitted by the interface excitons localized at lateral heterojunctions between TMD MLs is partially linearly polarized. Microscopically, this intrinsic polarization arises from the wave-vector dependent corrections to the optical matrix elements, which are related to the trigonal warping of electron and hole dispersions and the dependence of the effective masses on energy. We have calculated these corrections using the tight-binding model for a monolayer MoS$_2$. We have derived general analytical expressions for the Stokes parameters of the interface exciton emission valid for arbitrary crystallographic orientation of the interface. For the zigzag heterojunctions, optical transitions are polarized either along or perpendicular to the interface, whereas for the armchair interfaces, the direction of linear polarization is determined by the interplay of two mechanisms. In general, analysis of the degree and direction of linear polarization allows one to distinguish between the different types of heterointerfaces including the two different types of zigzag and armchair interfaces.

Using the variational approach, we have calculated interface exciton wave functions and energies in realistic lateral heterostructures taking into account an in-plane electric field applied perpendicular to the interface. Calculations show that the mechanism related to the dependence of the effective masses on energy is dominant, and the total degree of linear polarization of exciton emission can exceed 10\%. Finally, our results suggest that both the magnitude and the orientation of the linear polarization can be tuned by changing the applied electric field. 

\acknowledgements

We thank M. M. Glazov and S. A. Tarasenko for fruitful discussions.
We acknowledge financial support from the Russian Science Foundation Grant No. 25-72-10031 and the Foundation for the Advancement of Theoretical Physics and Mathematics ``BASIS''.

\appendix

\section{Tight-binding calculations of electron states and optical transitions}
\label{app:TB}

To support the results of the main text and calculate the key parameters $A$ and $\beta$ atomistically, we perform calculations of lateral heterostructures within the tight-binding (TB) model. We use the six-band TB model described, e.g., in Refs.~\cite{Rostami:2015,Silva-Guillen:2016,Ridolfi:2015,Rybkovskiy:2017aa}. The basis consists of $d_{z^2}$, $d_+$ and $d_-$ orbitals of the metal atom and $p_+$, $p_-$ and $p_{z,A}$ orbitals of the chalcogen atom. Here, $d_\pm = d_{x^2-y^2} \pm 2\i d_{xy}$, $p_\pm = p_{x,S} \pm \i p_{y,S}$, and $p_{\alpha, S} = (p_{\alpha,t} + p_{\alpha,b})/\sqrt{2}$, $p_{z, A} = (p_{z,t} - p_{z,b})/\sqrt{2}$ are the symmetric and asymmetric combinations of the $p$-type orbitals of the top ($t$) and bottom ($b$) chalcogen atoms in the unit cell.

The velocity operator in the TB model is calculated as~\cite{Ivchenko:2002}:
\begin{equation}
\label{velTB}
\bm v_{\alpha\alpha'} (\bm R, \bm R') = \frac{\i}{\hbar} (\bm R' - \bm R) h_{\alpha\alpha'}(\bm R, \bm R')\:, 
\end{equation}
where $\alpha$ denotes the atomic orbital, $\bm R = (X, Y)$ is the atom coordinate, and $h_{\alpha\alpha'}(\bm R, \bm R')$ are the hopping matrix elements. The interband velocity matrix element $\bm v_{cv} (\bm q)$ is then found by calculating the matrix elements of the operator~\eqref{velTB} between the valence-band and conduction-band states of the bulk monolayer with the following wave functions:
\begin{equation}
\psi_{n,\bm q} (\bm r) = \sum_{\alpha, \bm R} C_{n,\alpha} \e^{\i \bm q \cdot \bm R} \phi_{\alpha} (\bm r - \bm R)\:.
\end{equation}
Here, $\phi_{\alpha}(\bm r)$ are the wave functions of atomic orbitals, $n = c,v$ is the band index, and $C_{n, \alpha}$ are coefficients describing the Bloch amplitude. The resulting expression for $\bm v_{cv}$ is
\begin{multline}
\label{vcv_bulk_TB}
\bm v_{cv}(\bm q)  = \frac{\i}{\hbar}\sum_{\alpha, \bm R; \alpha', \bm R'} C_{c,\alpha}^* C_{v,\alpha'} \e^{\i \bm q \cdot (\bm R - \bm R')}  \\
\times (\bm R' - \bm R) h_{\alpha\alpha'} (\bm R, \bm R')\:.
\end{multline}

Comparison of the calculated $\bm v_{cv} (\bm q)$ in the vicinity of the $\bm K_\pm$-points with the phenomenological Eq.~\eqref{vcv_symmetry} allows us to find the parameters $\gamma_3$, $\alpha$, $\beta$ and $A$, see Tab.~\ref{tab1}. We perform this procedure for two sets of TB parameters corresponding to the MoS$_2$ monolayer taken from Refs.~\cite{Rybkovskiy:2017aa} and~\cite{Rostami:2015}. Note that both TB parametrizations result in similar values of the parameters $A$ and $\beta$. For the calculations throughout the paper, we use the first parametrization.

\begingroup
\renewcommand{\arraystretch}{1.5}
\begin{table}[!t]
\begin{center}
\begin{tabular}{p{1.7cm}  p{0.8cm}  p{0.8cm}  p{0.8cm} p{0.8cm} p{0.8cm} p{0.8cm} p{0.8cm}}
\hline
\hline
Parameters &  $E_g$ & $\gamma_3$ & $m_c$ & $m_v$ & $\alpha$ & $\beta$ & $A$ \\ 
\hline  
I  & 2.47 & 5.93 & 0.37 & 0.4 & 0.54 & 8.98 & 0.17  \\ 
II & 1.82 & 4.27 & 0.54 & 0.54 & 0.7 & 8.65 & 0.22 \\ 
\hline
\hline
\end{tabular}
\end{center}
\caption{\label{tab1} Band parameters of MoS$_2$ monolayer obtained from the tight-binding models of Ref.~\cite{Rybkovskiy:2017aa} (I) and Ref.~\cite{Rostami:2015} (II).
 The energy gap $E_g = E_c - E_v$ is given in eV, the units of $\gamma_3$ are eV$\cdot$\AA, the effective masses of conduction and valence bands $m_{c/v}$ are given in the units of $m_0$, the units of $A$ are \AA, the units of $\alpha$ and $\beta$ are \AA$^2$.
}
\end{table}
\endgroup

We further calculate linear polarization of optical transitions between single-particle electron and hole states localized at a lateral heterojunction. We consider zigzag interface (parallel to the $x$ axis, $\theta = \pi$) and model it using the following confining potential
\begin{equation}
\label{V_TB}
V(Y) = V_0 \Theta(Y) - e F Y
\end{equation}
with positive $V_0$ and $F$. Here, we do not take into account the Coulomb interaction between the electron and hole, so that the charge carriers are kept in the vicinity of the interface only by the applied electric field.

\begin{figure}[t]
\begin{center}
  \includegraphics[width=0.9\linewidth]{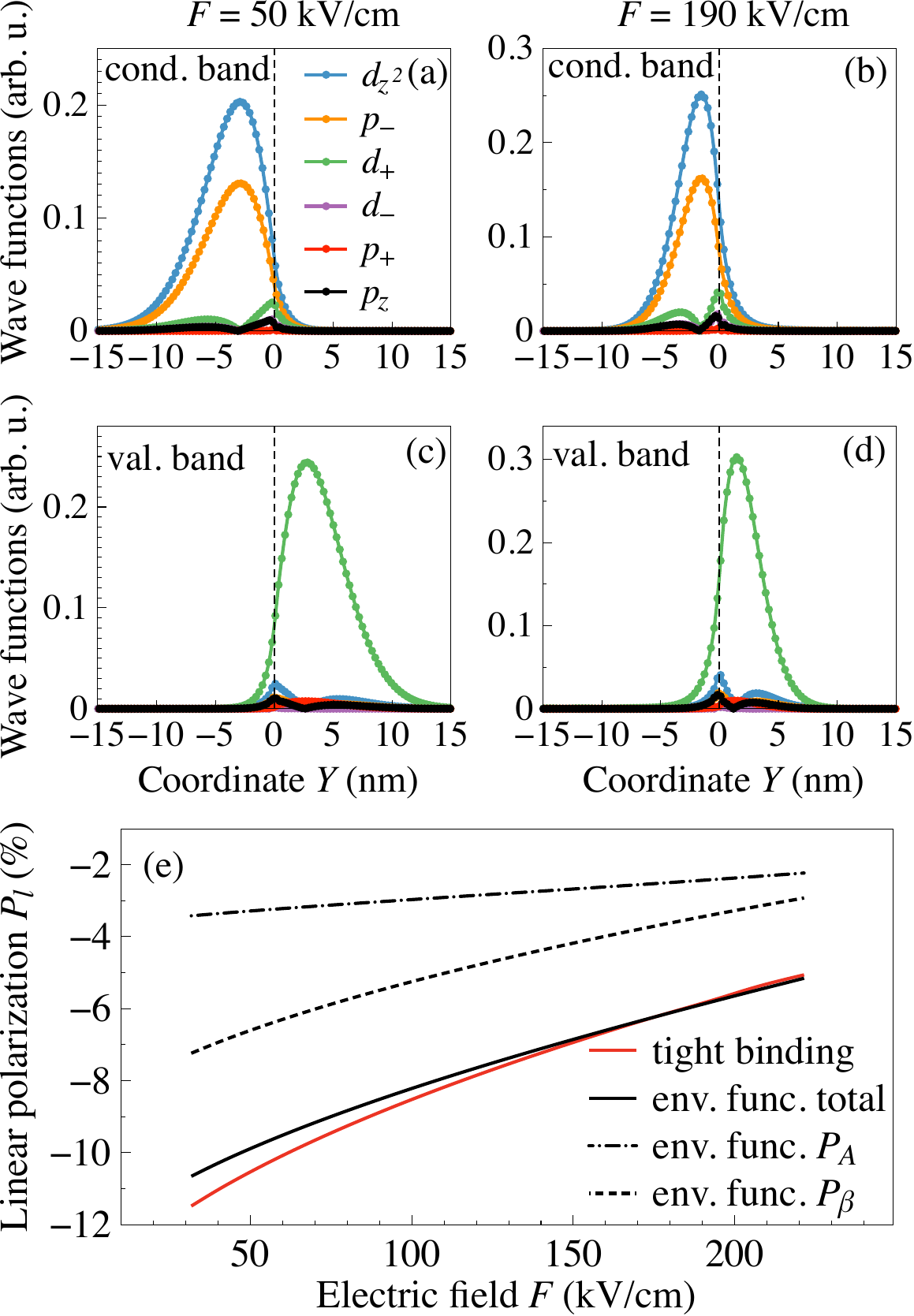}
 \end{center}
 \caption{\label{fig:Polariz_TB_kp} Tight-binding calculations of electron and hole states localized at the zigzag lateral heterojunction by applied electric field and linear polarization of corresponding optical transitions. (a-d) The absolute values of the coefficients $C_{n,Y,\alpha}$ in Eq.~\eqref{psi_zigzag}  describing contributions of different atomic orbitals in a wave function. The results are shown for the ground states of  (a, b) the conduction and (c, d) valence  bands in the potential Eq.~\eqref{V_TB} for $V_0 = 200$~meV and two values of electric field $F = 50$~kV/cm and $F = 190$~kV/cm. (e) Degree of linear polarization calculated using the TB model (red line) and the envelope function approximation (black line).  Contributions due to the trigonal warping ($P_A$) and the dependence of mass on energy ($P_\beta$) are shown by dashed-dotted and dashed lines, respectively.
  }
\end{figure}

The wave functions of the conduction and valence band states have the form
\begin{equation}
\label{psi_zigzag}
\psi_{n, q_x} (\bm r) = \sum_{\alpha,\bm R} C_{n, Y, \alpha} \e^{\i q_x X} \phi_{\alpha} (\bm r - \bm R)\:.
\end{equation}
The absolute values of the calculated coefficients $C_{n, Y, \alpha}$ are shown in Fig.~\ref{fig:Polariz_TB_kp}(a-d) for the ground states in the conduction and valence bands. We set $q_x = 4\pi/3a_0$, where $a_0$ is the lattice constant, which corresponds to the projection of the $\bm K_+$-point of the bulk Brillouin zone onto the interface. Just as in case of the $K_+$ valley in the bulk monolayer, the electron wave function is mainly formed by $d_{z^2}$ and $p_-$ atomic orbitals, whereas the hole wave function is determined by $d_+$ orbitals. The electron and hole reside mainly at the opposite sides of the heterojunction, but their wave functions weakly overlap in the vicinity of the interface. This is similar to the interface excitons bound by the Coulomb attraction, so one can apply a similar formalism to calculate linear polarization of optical transitions.

The symmetry of zigzag interface forces $P_{l'}=0$, while another Stokes parameter can be calculated as
\begin{equation}
P_l = \frac{|v_{cv, x}|^2 - |v_{cv, y}|^2}{|v_{cv, x}|^2+|v_{cv, y}|^2}\:,
\end{equation}
where $\bm v_{cv}$ is found by averaging the velocity operator~\eqref{velTB} between the wave functions $\psi_{v, q_x}$ and $\psi_{c, q_x}$ given by Eq.~\eqref{psi_zigzag} [analogous to Eq.~\eqref{vcv_bulk_TB}]. The calculated $P_l$ as a function of electric field is shown in Fig.~\ref{fig:Polariz_TB_kp}(e) by the red line. It is negative, so that the polarization is perpendicular to the interface. The absolute value of polarization degree decreases with increasing electric field.

This calculation can be directly compared to the results of the effective mass model of the same system (without Coulomb interaction). Solution of the Schr\"odinger equation with the potential~\eqref{V_TB} yields the following electron wave function:
\begin{equation}
\label{fcApp}
f_c (y) = \mathcal N \begin{cases}
 \mathrm{Bi}(-\lambda_c + v_c) \,\mathrm{Ai} \left(-y/l_c - \lambda_c \right),~y < 0\\
\mathrm{Ai}(-\lambda_c)\,\mathrm{Bi} \left(-y/l_c - \lambda_c + v_c \right),~y\geq 0\:,
\end{cases}
\end{equation}
where $\mathrm{Ai}$ and $\mathrm{Bi}$ are the Airy functions of the first and second kind, respectively, $l_c = (\hbar^2/2m_cF)^{1/3}$, $\lambda_c = \eps_c/(F l_c)$, $\eps_c$ is the electron energy found by the matching of wave functions at $y = 0$, and $v_c = V_0/(F l_c)$. The hole wave function $f_v$ is then found by replacing  $y \to -y$ and $m_c \to m_v$ in Eq.~\eqref{fcApp}.

The degree of linear polarization $P_l$ is then calculated from Eq.~\eqref{Xpol} for $\theta = \pi$, where $\kappa_1$ and $\kappa_2$ are found analogously to Eq.~\eqref{kappas} using $f_c(y)$ and $f_v(y)$ given by Eq.~\eqref{fcApp} with the effective masses $m_c$ and $m_v$ from Tab.~\ref{tab1}. This analytical approach allows us to separate contributions from the trigonal warping ($P_A=-A \kappa_1$) and the dependence of mass on energy ($P_\beta= -\beta \kappa_2$). These contributions are shown by the dashed-dotted and dashed lines in Fig.~\ref{fig:Polariz_TB_kp}(e), respectively. Just as in case of the calculations in the main text, Fig.~\ref{fig:X_Pl}, one can see that $|P_\beta|$ is larger then $|P_A|$. Their decrease with increase of $F$ is related to the increase of the overlap between electron and hole in larger fields, so that effectively smaller wave vectors contribute to the polarization.

The TB and envelope function calculations of $P_l$ agree particularly well at large electric fields~\footnote{The visible crossing of the curves in the field of about 175 kV/cm is apparently related to the fact that at large fields the wave functions start to significantly tunnel through barriers and oscillate.}.
This is because the latter calculation is perturbative and is valid exactly for the small values of $P_l$ only. Notably, even when the polarization is as large as 10\%, the difference between the envelope function and atomistic calculations remains smaller than 1\%. This validates our approach for calculation of the interface exciton linear polarization in the main text.

\renewcommand{\i}{\ifr}
\bibliographystyle{apsrev4-1-customized}
\bibliography{Linear_polarization_refs}

\end{document}